\documentclass[preprint,showpacs,showkeywords,preprintnumbers,amsmath,amssymb,superscriptaddress]{revtex4}
\usepackage{graphicx}
\usepackage{dcolumn}
\usepackage{bm}
\begin{document}

\title{Fractionally diffusing passing through the saddle point of metastable potential}

\author{Chun-Yang Wang\footnote{Corresponding author: wchy@mail.bnu.edu.cn}}
\affiliation{Shandong Provincial Key Laboratory of Laser Polarization and Information Technology, College of Physics and Engineering, Qufu Normal University, Qufu 273165, China}
\affiliation{State Key Laboratory of Theoretical Physics, Institute of Theoretical Physics, Chinese Academy of Sciences, Beijing 100190, China}
\author{Cui-Feng Sun}
\author{Hong Zhang}
\author{Xue-Mei Zong}
\affiliation{Shandong Provincial Key Laboratory of Laser Polarization and Information Technology, College of Physics and Engineering, Qufu Normal University, Qufu 273165, China}
\author{Ming Yi\footnote{Co-corresponding author: yiming@wipm.ac.cn}}
\affiliation{Key Laboratory of Magnetic Resonance in Biological Systems, Wuhan Institute of Physics and Mathematics, Chinese Academy of Sciences, Wuhan, China}


\begin{abstract}
The diffusion of a fractional Brownian particle passing over the saddle point is studied in the field of the metastable potential.
The barrier escaping probability is found to be greatly related to the fractional exponent $\alpha$.
Properties are revealed to move reversely in the opposite direction of diffusion when $\alpha$ is relatively large despite of the zero-approximating effective friction of the system.
This is very anomalous to the standard Brownian motion.
\end{abstract}



\maketitle

\section{INTRODUCTION}

Fractional Brownian motion (fBm) is a quintessential model for the stochastic processes which is characterized by long-memory stationary and Gauss-distributed increments \cite{fbm1,fbm2,fbm3}.
Different from the standard Brownian motion (sBm), fBm is generally believed to originate from the fractional Gaussian noise (fGn) \cite{fgn2,fgn1} whose autocorrelation is characterized by the Hurst exponent $H$.
For which, $H=1/2$ corresponds to the standard Brownian motion, $H<1/2$ and $H>1/2$ denote the sub- or super-diffusive cases, respectively.
Theoretically, no matter fBm or sBm, both of them can be categorized in the Langevin unification of fractional motions \cite{fms1} in the study of anomalous diffusions \cite{ad01,ad02,ad03}.

Since it offers an alternative model of random processes displaying the Joseph effect\cite{jef1}, fBm and its variants have attracted considerable attention in the past few decades \cite{fbm2,levy}.
Enormous efforts have been made on the establishment and development of relevant theories.
However, comprehensively review on the previous studies one can find, little information has been got concerning on the dynamical details of a fBm particle diffusing passing over the saddle point of a matastable potential energy surface (PES).
This maybe because of the fractional calculus difficulties immersed in the procedure of solving the kinetic differential equations.
But recently we find that this can be easily achieved from another road of mathematical deriving.
Therefore in this paper, we report our recent study on this point.

The paper is organized as follows:
in Sec.\ref{sec2}, kinematic relations of the diffusion particles are obtained by Laplacian solving the fractional Langevin equation.
The probability of successfully escaping from the potential well is computed in Sec.\ref{sec3} for a characteristic visualization.
Sec.\ref{sec4} serves as a short summary of present results where some discussions are also made for a further consideration.

\section{fractional Langevin equation and its solution}\label{sec2}

Let us begin from an alternative approach to fBm basing on the fractional generalized Langevin equation (GLE) \cite{fle1,fle2,fle3}
\begin{eqnarray}
m\frac{d^{2}x(t)}{dt^{2}}&=&F(x)-\bar{\gamma}\int^{t}_{0}(t-t')^{2H-2}\frac{dx(t)}{dt'}dt'+\xi^{H}(t)\nonumber\\
&=&F(x)-\bar{\gamma}\Gamma(2H-1)\frac{d^{2-2H}}{dt^{2-2H}}x(t)+\xi^{H}(t),
\label{gle}
\end{eqnarray}
where $\xi^{H}(t)$ is the fractional Gaussian noise (fGn) which is zero mean with autocorrelation $\langle\xi^{H}(t_{1})\xi^{H}(t_{2})\rangle=2D_{H}H(2H-1)|t_{1}-t_{2}|^{2H-2}$.
$D_{H}=[\Gamma(1-2H)\textrm{cos}(H\pi)]/(2H\pi)$ is the diffusion coefficient identified by the Gamma function $\Gamma(z)=\int_{0}^{\infty}t^{z-1}e^{-t}dt$.

Although Eq.(\ref{gle}) is the standard presentation of fBm, it is mathematically intractable due to the fractional calculus.
But fortunately we find in our recent study that it can be transformed to a more conventional form of fractional GLE by using of the Gamma function $\Gamma(s)$ and the relation between the external force $F(x)$ and the potential energy $U(x)$. Mathematically it can also be derived from a generalized system-plus-reservoir model of harmonic oscillators \cite{hotb}. i.e.
\begin{eqnarray}
m\frac{d^{2}x(t)}{dt^{2}}&=&F(x)-\bar{\gamma}\int^{t}_{0}\frac{1}{(t-t')^{\alpha}}\frac{dx(t)}{dt'}dt'+\xi(t)
\label{fgle}
\end{eqnarray}
and more conveniently
\begin{eqnarray}
m\ddot{x}+\int^{t}_{0}\eta(t-t')\dot{x}(t')dt'+\partial_{x}U(x)=\xi(t)
\label{fle}
\end{eqnarray}
where $\eta(t)=\eta_{\alpha}t^{-\alpha}/\Gamma(1-\alpha)$ is the frictional kernel with fractional exponent $0<\alpha<1$ and
$\eta_{\alpha}$ the strength constant.
$\xi(t)$ is the renewed fGn and $\langle\xi(t)\xi(t')\rangle=k_{B}T\eta_{\alpha}|t-t'|^{-\alpha}$ the fluctuation-dissipation theorem \cite{fdt1}.

The dominant advantage of Eq. (\ref{fle}) lies in its easy processing.
The solution of it, namely also the equation of motion for the diffusing particle, can be easily obtained by a series of Laplacian transformation. For an example, in the particular case of an inverse harmonic potential $U(x)=-\frac{1}{2}m\omega^{2}x^{2}$, we find after some algebra
\begin{eqnarray}
x(t)=\langle x(t)\rangle+\int^{t}_{0}H(t-\tau)\xi(\tau)d\tau,
\end{eqnarray}
in which the mean position of the particle along the transport direction is given by
\begin{eqnarray}
\langle x(t)\rangle=[1+\omega^{2}\int^{t}_{0}H(\tau)d\tau]x_{0}+H(t)v_{0},
\end{eqnarray}
and the variance of $x(t)$ reads
\begin{eqnarray}
\sigma^{2}_{x}(t)=\int^{t}_{0}dt_{1}H(t-t_{1})\int^{t_{1}}_{0}dt_{2}\langle
\xi(t_{1})\xi(t_{2})\rangle H(t-t_{2}),
\end{eqnarray}
where $H(t)=\mathcal{L}^{-1}[(s^{2}+s\eta(s)-\omega^{2})^{-1}]$ is namely the response function and $\eta(s)=\eta_{\alpha}s^{\alpha-1}$ is the Laplacian transformation of the friction kernel $\eta(t)$.

\section{diffusion dynamics of fractional damping system}\label{sec3}

In the studies here and following, we extend the results aforesaid to the more realistic case of a matestable potential to reveal the diffusion dynamics of a fractional damping (FD) system described by Eq. (\ref{fle}) \cite{chyx}.
This means to set
\begin{eqnarray}
     U(x)=\left\{
     \begin{array}{ll}
     U_{b}-\frac{1}{2}m\omega^{2}_{b}(x-x_{b})^{2}\quad & \textrm{near barrier} ;\\
     \frac{1}{2}m\omega^{2}_{a}(x-x_{a})^{2}\quad & \textrm{near well}.
     \end{array}
     \right.
\end{eqnarray}
where $U_{b}$ denotes the barrier height and $\omega_{i}$ $(i=a,b)$ indicates the frequency near the potential well as well as the saddle point.
Although some new derivations should be made, but noticing that $U(x)$ can be approximated to be an inverse harmonic potential in the neighbourhood of the saddle point, this would not be much of a problem.

For the study of diffusion dynamics, one of the basic task is to compute the successful rate of the particle escaping from the potential well.
Mathematically, it always emerges to be a complementary error function due to the Gaussian property of noise $\xi(t)$ and linearity of the GLE. i.e.
\begin{eqnarray}
P(t)=\int^{\infty}_{0}W(x,t)dx=\frac{1}{2}\textrm{erfc}\left(-\frac{\langle
x(t)\rangle}{\sqrt{2A_{11}(t)}}\right),\label{eq,chsi}
\end{eqnarray}
yielding from a short integration on the reduced distribution function $W(x,t)$ where
\begin{eqnarray}
W(x,t) =\frac{1}{\sqrt{2\pi A_{11}(t)}}\textrm{exp}\left[{-\frac{(x-\langle x(t)\rangle)^{2}}{2A_{11}(t)}}\right], \label{eq,pd}
\end{eqnarray}
is resulted from the joint probability density function (PDF) of the system \cite{adelm}
\begin{eqnarray}
W(x,v,t)
=\frac{1}{2\pi|\textbf{A}(t)|^{1/2}}e^{-\frac{1}{2}\left[y^{\dag}(t)
\textbf{A}^{-1}(t)y(t)\right]},\label{eq,pdf}
\end{eqnarray}
with $y(t)$ the vector $[x-\langle x(t)\rangle,v-\langle
v(t)\rangle]$ and $\textbf{A}(t)$ the matrix of second moments
\begin{subequations}
\begin{eqnarray}
A_{11}(t)&=&\sigma^{2}_{x}(t)=\langle[x-\langle x(t)\rangle]^{2}\rangle,\\
A_{12}(t)&=&A_{21}(t)=\langle[x-\langle x(t)\rangle][v-\langle
v(t)\rangle]\rangle,\\
A_{22}(t)&=&\sigma^{2}_{v}(t)=\langle[v-\langle v(t)\rangle]^{2}\rangle.
\end{eqnarray}\label{A(t)}\end{subequations}
Now everything is ready before tracking the trajectories to find the dynamical details.

\begin{figure}[ht]
\centering
\includegraphics[scale=0.9]{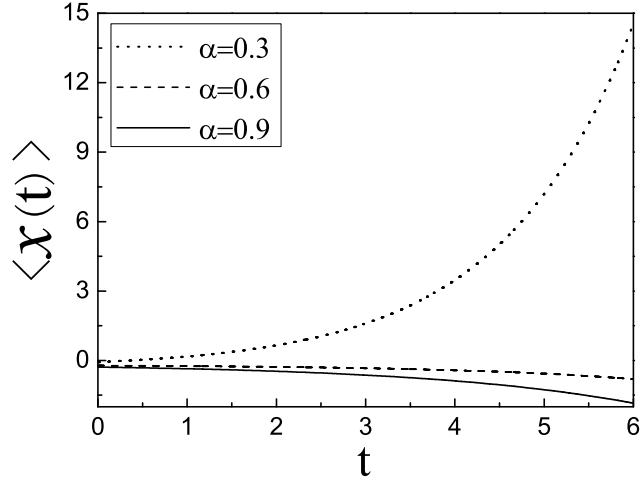}
\caption{Time dependent varying of $\langle x(t)\rangle$ at various $\alpha$. Parameters in use are
$\eta_{\alpha}=2.0$, $\omega=k_{B}T=1.0$, $x_{0}=-1.0$ and $v_{0}=2.0$. \label{fig1}}
\end{figure}

\begin{figure}[ht]
\centering
\includegraphics[scale=0.9]{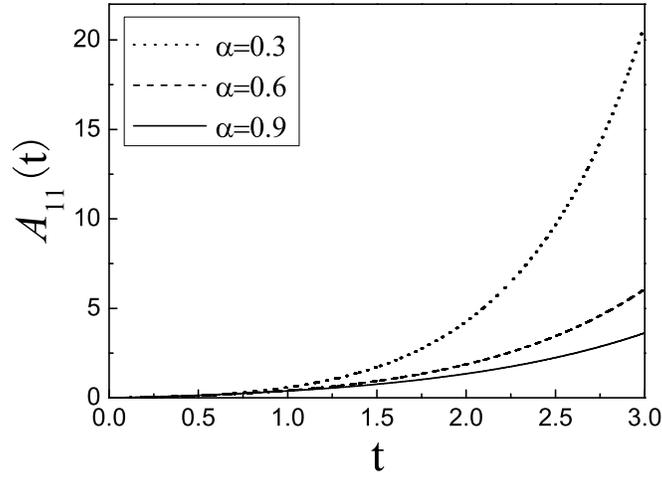}
\caption{Time dependent varying of $A_{11}(t)$ at various $\alpha$. \label{fig2}}
\end{figure}

In the calculations here and following, we rescale all the quantities so that dimensionless unit such as $k_{B}T=1.0$ is used.
Firstly in Figs. \ref{fig1} and \ref{fig2}, we plot the time dependent varying of $\langle x(t)\rangle$ and $A_{11}(t)$ (namely also $\sigma^{2}_{x}(t)$) at various $\alpha$.
From which we can see that $\langle x(t)\rangle$ increases quickly when $\alpha$ is relatively small. But as the increasing of $\alpha$,
$\langle x(t)\rangle$ suddenly decays into a negative diverging function of $\alpha$.
The critical point lies in the nearest neighbourhood of $\alpha=0.55$ in the particular case of what is considered here.
This is a very nontrivial result, because from the view point of PDF evolution, $\langle x(t)\rangle$ indicates the center of the wave packets and $A_{11}(t)$ the width.
Therefore it reveals that the center of the wave packet may move in the opposite direction of diffusion.

\begin{figure}[ht]
\centering
\includegraphics[scale=0.9]{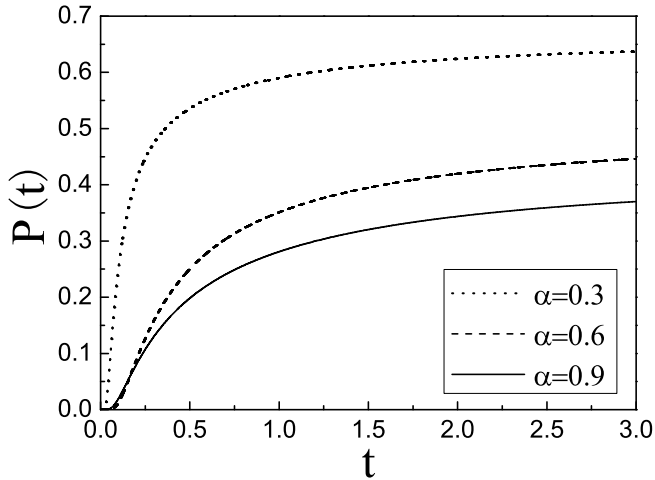}
\caption{Time dependent varying of $P(t)$ at various $\alpha$. \label{fig3}}
\end{figure}

However, the value of $A_{11}(t)$ is always positive increasing.
This means, as the forwarding of the center the width the wave packet is always expanding.
Therefore, despite of the unusual movement of the center one may always expect a steady barrier escaping probability.
As is shown in Fig. \ref{fig3}, the probability for a particle to diffusing passing the saddle point tends to be a steady one (defined as $P_{\textrm{st}}$) in the long time limit no matter what is the value of $\alpha$.
The occurrence of such a nontrivial result may probably be caused by the property of long-range correlations (the ``Joseph effect'') immersed in the standard process of fBm \cite{fbm2}.

\begin{figure}[ht]
\centering
\includegraphics[scale=0.9]{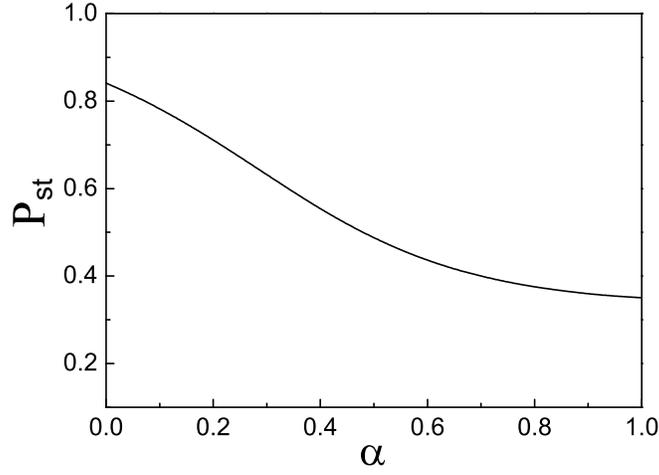}
\caption{Stationary value $P_{\textrm{st}}$ plotted as a function of $\alpha$.  \label{fig4}}
\end{figure}

\begin{figure}[ht]
\centering
\includegraphics[scale=0.9]{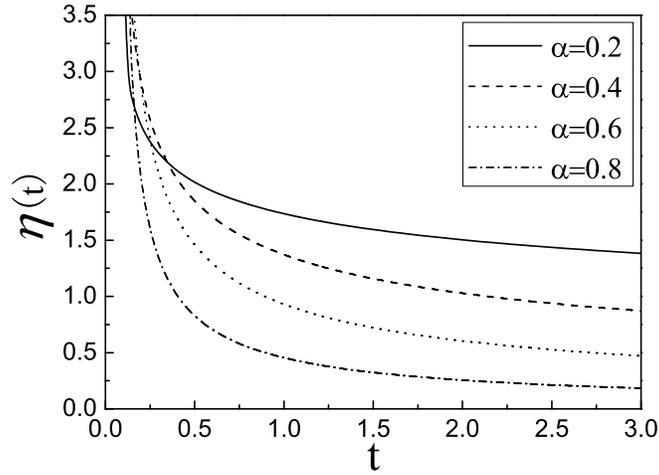}
\caption{Friction kernel $\eta(t)$ plotted as a function of $t$ at various $\alpha$.  \label{fig5}}
\end{figure}

In order to get more details about the fBm, we plot in Fig. \ref{fig4} the time dependent varying of $P_{\textrm{st}}$ (the steady value of $P(t)$) as a function of the fractional exponent $\alpha$. From which we can see that $P_{\textrm{st}}$ decays monotonously as the increasing of $\alpha$. At first thought, this is considered to be caused by a strong resistance from the friction kernel. However, when we check through the time-dependent varying of $\eta(t)$, a startling result is found that the effective friction of the system tends to be zero as the increasing of $\alpha$. As is shown in Fig. \ref{fig5} for more explicitness. But what is the real reason then?
Zero damping will lead to low barrier escaping, right?

\begin{figure}[ht]
\centering
\includegraphics[scale=0.9]{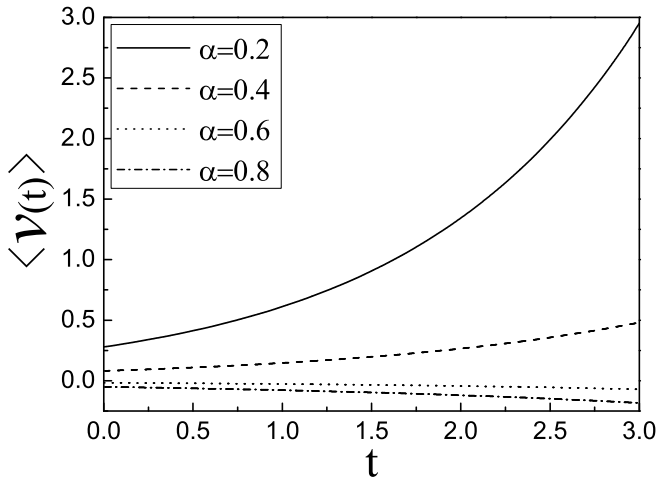}
\caption{Friction kernel $\eta(t)$ plotted as a function of $t$ at various $\alpha$.  \label{fig6}}
\end{figure}

In order to answer these questions, we make a thorough investigation on the kinematical properties of the system.
We find that the mean velocity of the particle can be obtained from a short derivation from $\langle x(t)\rangle$, i.e.
\begin{eqnarray}
\langle v(t)\rangle=\frac{d}{dt} \langle x(t)\rangle=H(t)x_{0}+\frac{d}{dt}H(t)v_{0}.
\end{eqnarray}
And succeeding calculation reveals that $\langle v(t)\rangle$ tends to be positive when $\alpha$ is relatively small, but will turn to be negative at large $\alpha$.
As is shown in Fig. \ref{fig6} for more details.
The critical point is identical to what we have found hereinbefore.
Therefore we have known the reasons now:
for the fBm process with large $\alpha$, the effective friction of the system is very strong at the beginning.
Particle inserted in this dissipative environment loses its initial kinetic energy quickly. After a very short period of time, it gets a reversed kinetic energy
to move in the opposite direction of diffusion despite of the zero-approximating damping.
This results in a low probability of barrier escaping (small $P_{\textrm{st}}$) in the long time.

\section{summary and discussion}\label{sec4}

In conclusion, we have studied in this paper the diffusion dynamical properties of the fractional damping system which may induce fBm.
Several anomalous phenomena are revealed such as the reverse moving of the particles at large $\alpha$, the low $P_{\textrm{st}}$ at zero-approximating effective friction and so on.
All these results have shown us a vivid picture much different from the standard Brownian motion.

Since theoretically fBm may be a model particularly relevant to subcellular transport, unbiased translocation \cite{sd26,sd27}, the dispersion of apoferritin proteins in crowded dextran solutions \cite{sd11} and lipid molecules in lipid bilayer \cite{sd12}.
Therefore we believe that deep research will be greatly indeed in the near future in some disciplinary field such as sub-diffusing mRNA molecules \cite{sd10}, RNA proteins and chromosomal loci within E. coli cells \cite{sd04}.
And the results we have got here will in no doubt motivate the continuous demystification of these seemingly simple subjects.

\section * {ACKNOWLEDGEMENTS}

This work was supported by the Shandong Province Science Foundation for Youths under the Grant No.ZR2011AQ016, the Open Project Program of State Key
Laboratory of Theoretical Physics, Institute of Theoretical Physics, Chinese Academy of Sciences, China under the Grant No.Y4KF151CJ1, the National Natural Science Foundation of China under the Grant No.11275259 and 91330113.

\end{document}